\documentclass[useAMS,usenatbib]{mnras}

\usepackage{graphicx}
\usepackage[dvipsnames]{color}
\usepackage[normalem]{ulem}
\usepackage{threeparttable}
\usepackage{multirow}
\usepackage{array}

\usepackage{etoolbox}
\makeatletter
\newcount\c@additionalboxlevel
\newcount\c@maxboxlevel
\patchcmd\@combinedblfloats{\box\@outputbox}{%
    \stepcounter{additionalboxlevel}%
    \box\@outputbox
}{}{\errmessage{\noexpand\@combinedblfloats could not be patched}}

\AtBeginShipout{%
    \ifnum\value{additionalboxlevel}>\value{maxboxlevel}%
    \typeout{Warning: maxboxlevel might be too small, increase to %
        \the\value{additionalboxlevel}%
    }%
    \fi
    \@whilenum\value{additionalboxlevel}<\value{maxboxlevel}\do{%
        \typeout{* Additional boxing of page `\thepage'}%
        \setbox\AtBeginShipoutBox=\hbox{\copy\AtBeginShipoutBox}%
        \stepcounter{additionalboxlevel}%
    }%
    \setcounter{additionalboxlevel}{0}%
}
\makeatother

\newcommand{\NII}{[N~{\sc ii}]\ }
\newcommand{\HII}{H~{\sc ii}\ }

\newcommand{\Ha}{H$\alpha$\ }
\newcommand{\kms}{\,\mbox{km}\,\mbox{s}^{-1}}

\newcommand{\SIIHa}{I([S~{\sc ii}])/I(H$\alpha$)}

\newcommand{\NIIHa}{I([N~{\sc ii}])/I(H$\alpha$)}
\newcommand{\OIIIHb}{I([O~{\sc iii}]5007)/I(H$\beta$)}
\newcommand{\be}{\begin{equation}}
\newcommand{\ee}{\end{equation}}

\def \gtsima{$\, \buildrel > \over \sim \,$}
\def \ltsima{$\, \buildrel < \over \sim \,$}
\def \simgt{\lower.5ex\hbox{\gtsima}}
\def \simlt{\lower.5ex\hbox{\ltsima}}

\begin{document}

\title[Metallicity and ionization in PRGs.] {Metallicity and ionization state of the gas in polar-ring galaxies}

\author[Egorov et al.]{
    Oleg V.~Egorov$^{1,2}$\thanks{E-mail: egorov@sai.msu.ru} and Alexei V.~Moiseev$^{1,2}$\thanks{E-mail: moisav@sao.ru}
    \\
    $^{1}$ Lomonosov Moscow State University, Sternberg Astronomical Institute,
    Universitetsky pr. 13, Moscow 119234, Russia
    \\
    $^{2}$ Special Astrophysical Observatory, Russian Academy of Sciences, Nizhny Arkhyz 369167, Russia
    \\
}

\date{Accepted 2019 Month 00. Received 2019 Month 00; in original
    form 2019 Month 00}

\pagerange{\pageref{firstpage}--\pageref{lastpage}} \pubyear{2019}

\maketitle

\label{firstpage}

\begin{abstract}

The ionization state and oxygen abundance distribution in a sample of polar-ring galaxies (PRGs) were studied from the long-slit spectroscopic observations carried out with the SCORPIO-2 focal reducer at the Russian 6-m telescope. The sample consists of 15 PRGs classified as  `the best candidates' in the  SDSS-based Polar Ring Catalogue. The distributions of line-of-sight velocities of stellar and gaseous components  have given kinematic confirmation of polar structures in 13 galaxies in the sample. We show that ionization by young stars dominates in the external parts of polar discs, while shocks have a significant contribution to gas excitation in the inner parts of polar structures. This picture was predicted earlier in a toy model implying the collision between gaseous clouds on polar orbits with the stellar disc gravitational potential well. The exception is a moderately inclined ring to the host galaxy NGC 5014: the accreted gas in the centre has already settled on the main plane and ionized by young stars, while the gas in the internal part of the ring is excited by shocks. The present study three times increases the number of polar structures  with an available  oxygen abundance estimation. The measured values of the gas metallicity  almost do not depend on the galaxy luminosity. The radial [O/H] gradient in the considered polar rings is shallow or absent. No metal-poor gas was detected. We ruled out the scenario of the formation of polar rings  due to cold accretion from cosmic filaments for the considered sample of PRGs.

\end{abstract}


\section{Introduction}
\label{sec:intro}

\begin{figure*}
	\includegraphics[width=0.95\linewidth]{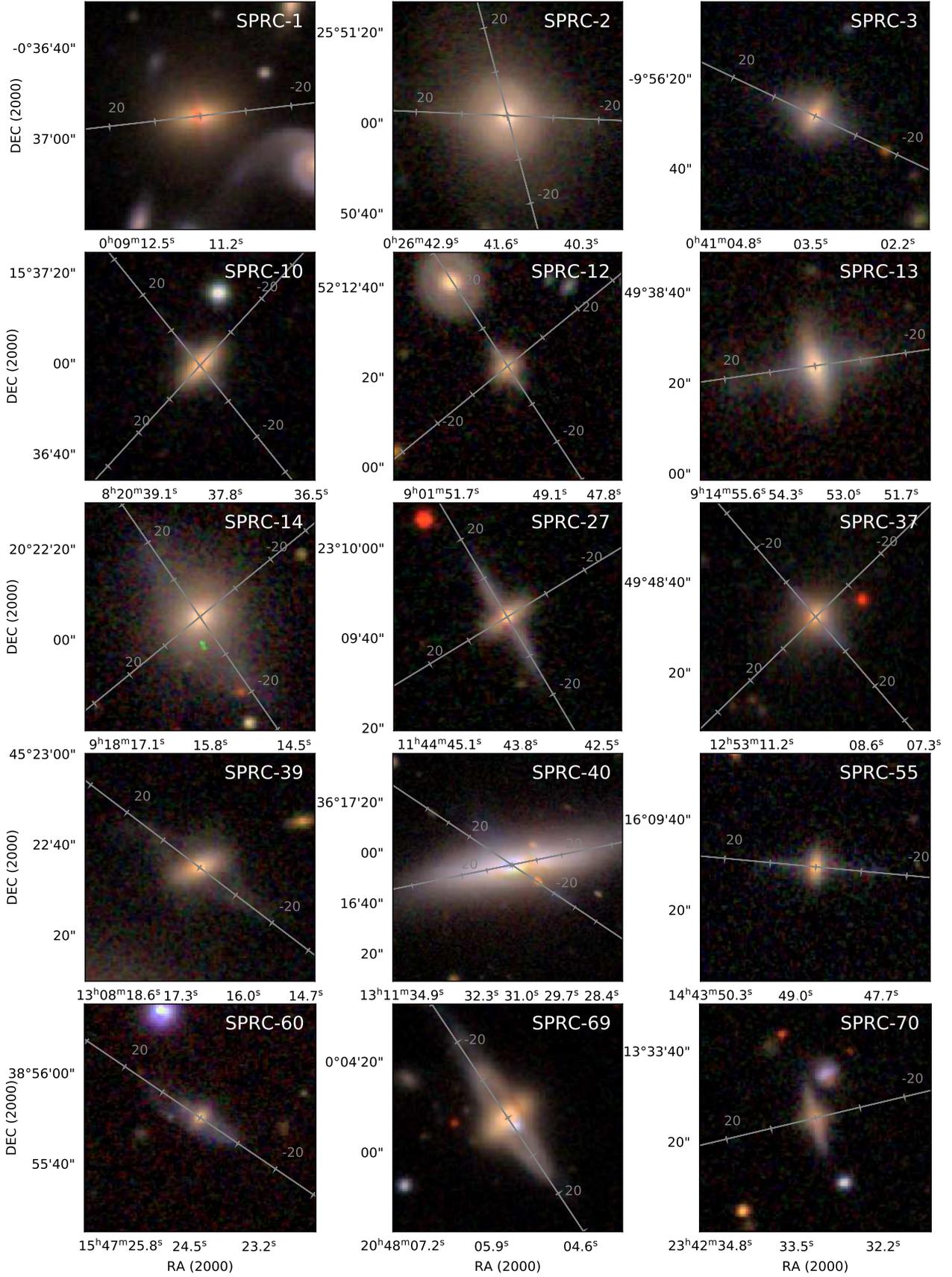}
	\caption{Colour SDSS DR12 images of the sample galaxies in the $g,r,i$ filters. The SCORPIO-2 slit positions are  overlapped.}\label{fig:sdss}
\end{figure*}

\begin{table*}
  \caption{Log of the observed data and basic information about the galaxies}
	\label{tab:data_info}
  \begin{threeparttable}
	\begin{small}
		\begin{tabular}{lrrrllll}
			\hline
			Galaxy  &  cz$^a$, $\kms$ & $R_{eff}^b$, \arcsec & Slit PA, deg  & Date of obs & $\mathrm{T_{exp}}$, s & $\theta$, \arcsec &  $\Delta\lambda$, \AA  \\
			\hline
			SPRC-1   & 21960  & 6.7$^c$ & 97  & 03/04 Dec 2013 & 4500 & 1.7 & 3700--7250 \\
			{SPRC-2}&		10523$^d$ & 	6.9 & 15  & 10/11 Oct 2012 & 4800 & 2.4 & 3700--7250 \\
			& & & 88 & 22/23 Sep 2011 & 7200 & 1.9 &3700--7250 \\
			SPRC-3 &  11063 &   2.8 & 65   & 11/12 Oct 2012 & 6000 & 1.2 & 5010--7310  \\
			{SPRC-10} & 12736  & 2.9 & 39  & 07/08 Dec 2010 & 4800  & 1.7 & 3700--7250  \\
			& & & 138 & 07/08 Dec 2010 & 6000 & 1.5 &3700--7250 \\
			{SPRC-12} &	18728 &  1.9 &  33 & 22/23 Dec 2011 & 3600 & 1.2 & 3700--7250  \\
			& 						&  											 &  310 & 22/23 Dec 2011  & 4800 & 1.3 & 3700--7250 \\
			SPRC-13 &9522 &   3.1 &  98 & 19/20 Oct 2014 & 5400  & 2.8 & 3700--7250  \\
			SPRC-14 & 9548  &   4.4 &  35 & 08/09 Dec 2010 & 3600 & 2.4 & 3700--7250  \\
            &   &  &  130 & 08/09 Dec 2010 & 3600 & 2.4 & 3700--7250  \\
			{SPRC-27} & 14509   &   2.1 &  31 & 20/21 Feb 2012 & 4800 & 1.0 & 3700--8530   \\
			& 		 								&       					 &  122 & 20/21 Feb 2012 & 7200 & 1.2 & 3700--7250  \\
			{SPRC-37} & 20297  &   8.3 &   220& 26/27 Mar 2014& 4500  & 1.0& 3700--7250 \\
			& & & 134 & 26/27 Mar 2014& 2700  & 1.0& 3700--7250 \\
			SPRC-39 & 8792   &  4.5 &   53& 08/09 Dec 2014  & 4500 &2.2 & 3700--7250 \\
			SPRC-40 & 1126   &      13.8$^c$ &   102& 27/28 Feb 2014 & 4500 & 2.6&3700--7250  \\
			& 		 								&       					 &  56 & 14/15 Feb 2019 & 4800 & 1.8 & 3700--7250  \\
			SPRC-55 & 25782    &     1.6 &    85&  31 Mar/01 Apr 2014 & 2700 & 2.9 & 3700--7250  \\
			SPRC-60 &  23519 & 4.6$^c$ & 56 & 11/12 Sep 2010 & 4800 & 1.2 & 4800--8500 \\
			SPRC-69 & 7396   &   2.1&   214& 12/13 Nov 2014 & 5400 & 1.5 & 3700--7250   \\
			SPRC-70 & 20595 & 2.4 & 104 & 23/24 Sep 2011 & 4800 & 1.9 &  3700--7250 \\
			\hline
		\end{tabular}
		
	\begin{tablenotes}
      \item $^a$ Taken from \citet{sprc}
	  \item $^b$ Calculated from \citet{Reshetnikov2015}
	  \item $^c$ Taken from \citet{DR12} through the Hyperleda database
	  \item $^d$ Estimated in this work
    \end{tablenotes}
	\end{small}
  \end{threeparttable}
\end{table*}

Polar-ring galaxies (PRGs) are the most famous type of so-called multi-spin galaxies \citep[the term was introduced by][]{Rubin1994} which represent the systems with an external disc or ring rotating at the plane almost perpendicular to the host galaxy plane. Their unusual morphology allows us to investigate a wide range of issues related to their galactic formation and evolution: the baryonic matter accretion, the rate of galactic interaction, the 3D distribution of mass in their dark halos, etc.     

The first catalogue of PRGs candidates selected from photographic atlases was compiled by \citet{Whitmore1990}. It contained 157 objects including six kinamatically-confirmed polar rings (i.e., the rotation with the same systemic velocity was detected in two almost orthogonal planes), and 27 galaxies were included in the category of `good' candidates. The next generation catalogue: the SDSS-based Polar Rings Catalogue (SPRC) was presented by \citet{sprc}. It contains 275 objects, 185 among  them were classified as `good' and `best' candidates identified in the Sloan Digital Sky Survey (SDSS). A recent paper by \citet{Reshetnikov2019} { added 31 new possible  PRGs} found in SDSS. Thus, currently we know several hundreds PRGs candidates; several tens of them are kinematically confirmed \citep*[see details and references in ][]{Reshetnikov2004,sprc}.

Despite a large amount of known PRGs, they are still quite poorly studied. In particular, the mechanism of formation of polar components in PRGs is a topic of hot debates. Several scenarios were proposed: a major merging  of two orthogonally orientated disc galaxies \citep{Bekki1997, Bournaud2003}; a tidal accretion from a disrupted small companion on a polar orbit or from a gas-rich donor galaxy \citep{Schweizer1983, Reshetnikov1997}; a cold accretion from   cosmic filaments that are inclined to the host galaxy disc \citep{Maccio2006}. The last one seems to be the most intriguing among them. It is believed that the gas accretion along filaments is a very common phenomenon in a galaxy mass assembly at high and even at low redshifts \citep{Dekel2009}, but due to chemical and dynamical evolution of galaxies it is very difficult to find traces of this process \citep[see][for review]{SanchezAlmeida2014}. \citet{Steiman1982} showed that polar orbits around the host galaxy in the case of oblate or triaxial gravitational potential are very stable and slowly evolved.  If the cold accretion is indeed responsible for formation of some fraction of  polar rings, then PRGs should still represent  footprints of this process. The important clue to distinguish between different scenarios listed above should be measurements of chemical abundances of the PRGs.

The most studied nearby galaxy NGC~4650A \hbox{(PRC A-1)} is now considered as one of the best  candidates experienced the cold cosmic filament accretion. The extended and broad polar ring  \citep[or `a polar disc' according to][]{Iodice2006} possesses a significant fraction of the total baryon mass in the system, while a relatively low stellar velocity dispersion of the lenticular host contradicts to the major merger formation  of the polar component. The gas metallicity of the  polar disc ($Z\approx0.2Z_\odot$)  is lower than that expected for ordinary spirals of similar luminosity \citep{Spavone2010,Iodice2015}. The observed properties of this galaxy are reproduced in  simulations of the gas accretion along the filament \citep{Maccio2006,Brook2008,Snaith2012}. 

It is not clear whether the case of NGC~4650A is common or unique. For instance, the SPRC-7 galaxy has a  similar morphology, i.e., a broad relatively massive polar disc around the dynamically cold central S0 hosts, and  could be considered as a distant  counterpart of NGC 4650A \citep{Brosch2010,Khoperskov2014}. Whereas the gas oxygen abundance in the SPRC-7 polar disc is not very low \citep{Brosch2010}.

Up to now, the metallicity measurements were made only for few PRGs. For some of them, the authors actually obtained low gas metallicity values which are consistent with the hypothesis of a cold accretion from a cosmic filament like that in NGC~4650A \citep*[see, e.g.,][and references therein]{Spavone2015}. Unfortunately, the majority of  the available results demonstrate large uncertainties exceeding 0.3--0.5 dex \citep[see Fig.~9 in][]{Spavone2015}.
Therefore, new accurate estimations of gas metallicity in a large sample of PRGs are still needed to distinguish  between possible evolution scenarios of their formation.

The measurements of metallicity in polar or  strongly inclined gaseous discs might be biased because of the expected contribution of shock ionization. \citet{Wakamatsu1993} proposed that the shocks can be generated when the gas on polar orbits crosses the potential well of a stellar disc (like the galactic shock waves in spiral arms and bars), but  still there are no observations  supporting this hypothesis among  PRGs. The  exception is  a couple of galaxies NGC~660 (PRC C-13) and Arp~212 (PRC D-45), where the direct collision between gaseous components in the main and polar (inclined) discs are confirmed \citep{Moiseev2014late}. Shocks are also observed in some lenticular  galaxies that captured the kinematically misaligned external  gas  in considerably inclined planes \citep{Silchenko2009, Katkov2011, Katkov2014}, but this issue almost had not been tested in classical PRGs with relatively massive stellar-gaseous non-complanar components.

In order to increase the number  of PRGs with known  dynamical and chemical abundance properties, we  performed long-slit optical spectroscopic observations of the targets from the SPRC catalogue at the 6-m telescope of the Special Astrophysical Observatory of the Russian Academy of Sciences (SAO RAS). In this paper,  we present results of the study of gas ionization conditions and metallicities in 15 galaxies  having the emission lines bright  enough for the standard spectrophotometrical analysis. 
The paper is organized as follows: Section~\ref{sec:obs} describes the performed observations and data reduction. Sections~\ref{sec:kin}, \ref{sec:ionization}, and \ref{sec:metallicity}   present the results on the kinematics, ionization state, and gas metallicity, respectively. In Section~\ref{sec:discus}, we consider the luminosity--metallicity relation and discuss the obtained results in terms of the evolution scenarios of PRGs.

\section{Observations and data analysis}
\label{sec:obs}

\begin{figure*}
	\includegraphics[width=\linewidth]{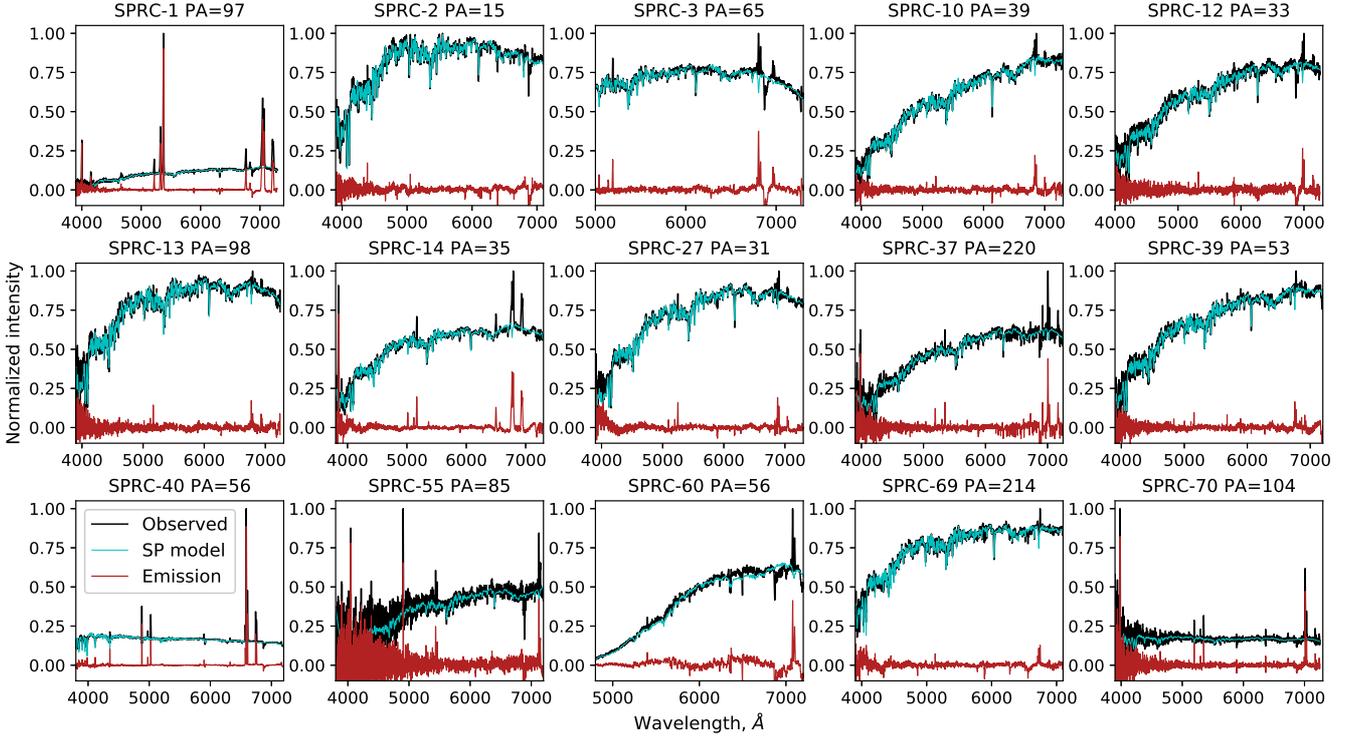}
	\caption{Examples of the observed spectra of the central parts of the sample galaxies (black) together with the modelled spectra of the stellar population (cyan) and with the residual emission spectrum of the ionized gas (red).}\label{fig:spectra}
\end{figure*}

The SPRC gives  an opportunity to analyse the structural properties of PRGs using the uniform SDSS photometric data. However, the spatially-resolved spectroscopic observations are essential for kinematic confirmation  of polar structures to study the dynamical properties of the gas and stars,  their ionization conditions, and chemical abundance. For this purpose in 2010--2019 we observed about 30 SPRC galaxies   mostly selected from the `best' or `possible face-on rings' categories \citep[see the classification in][]{sprc}.
The analysis of stellar and gas kinematics of the observed galaxies will be considered in the forthcoming paper, whereas  this paper presents the spectrophotometric data for 15 galaxies,  in which the detected emission lines were bright enough for measuring the ionized gas chemical abundance and excitation properties.

The observations  were made at the prime focus of the 6-m telescope of SAO RAS using the \mbox{SCORPIO-2} multi-mode focal reducer \citep{scorpio2} operating in the long-slit spectrograph mode with the 6~arcmin $\times$ 1~arcsec slit. The scale along the slit was 0.36~arcsec
per pixel. The total exposure time varied from 2700 to 7200 sec and the seeing -- from 1.0 to 2.9 arcsec for different objects. For seven of the observed targets, we obtained the spectra at two slit positions that crossed both main and polar discs; for the rest galaxies, the slit was aligned along the polar rings (see Fig.~\ref{fig:sdss}) or along the host disc (in the case of SPRC-1 and SPRC-70).

Table~\ref{tab:data_info} gives the log of observations and summarizes  the main parameters of galaxies. It shows the name of each galaxy according to the SPRC catalogue; its redshift; effective radius ($R_{eff}$) in r-SDSS band; position angle of the spectrograph slit ($PA$); date of observation; total exposure ($T_{exp}$);  seeing ($\theta$); spectral range ($\Delta\lambda$). 

Observations of the most galaxies were performed with the VPHG1200@540 grism providing the spectral resolution $\delta\lambda = 4.5-5.4$~\AA\, estimated as the FWHM of air glow lines. SPRC-3 and SPRC-27 (PA$=31\degr$) were observed with   the VPHG940@640 grism of a lower resolution ($\delta\lambda \simeq 7$~\AA).  The CCD detector  was \hbox{E2V 42-90} $4.5K\times 2K$ during observations of all the galaxies except for  SPRC-3 which was observed with the detector of a smaller format, $2K\times 2K$ CCD \hbox{EEV 42-40} .

The initial data reduction was performed in a standard way using the \textsc{idl} software package written for reducing long-slit spectroscopic data obtained with \mbox{SCORPIO-2} as it was described in our previous papers \citep[e.g.,][]{Egorov2018}. After the bias subtraction and cosmic rays hits removal, all individual exposures of each spectrum were stacked; further steps of data reduction include  correction of the geometric distortion, normalization to the flat-field spectrum, wavelength calibration, subtraction of the night-sky spectrum, and flux calibration using one of the spectrophotometric standards observed in the same night in a  close airmass with the galaxy. 

The spectrum of the SPRC-60 was not corrected for spectral sensitivity and, hence, was not flux-calibrated because of the absence of the standard star observations in the corresponding night. Further results on this galaxy are based only on the neighbouring line ratios.

Before analysing the emission lines, the stellar population models were subtracted from the spectra. These models were recovered from the observed spectra using the \textsc{ULySS} package by \cite{ulyss} and the grid of P{\'E}GASE-HR models (their validation was tested in \citealt{ulyss_models}). Fig.~\ref{fig:spectra} demonstrates the example of the spectra of central region of each observed galaxy together with the results of its decomposition to the spectra of the stellar population and of the ionized gas.

In order to increase the signal-to-noise ratio, all the spectra used for the \textsc{ULySS} analysis were binned along the slit using the 3 px ($\sim1''$) bins in the central region and 5 px (1\farcs8) bins at the distances $r>5''$ from the nucleus. The analysis of the ionized gas emission was performed with the original binning (1 px = 0\farcs36).  

The emission lines were fitted by a single Gaussian  in each spatial bin along the slit using the \textsc{idl} \textsc{mpfit} library \citep{mpfit}. The obtained fluxes of the emission lines were used to analyse the ionization condition and gas metallicity. 

In order to compare the chemical abundance values with the integral  luminosity of the galaxies, we made an aperture photometry of SDSS DR12 \citep{sdss_dr12} images in the $g$ and $r$ bands. The integration was performed in the polygonal apertures included both the galaxy and polar structures without the foreground star contamination. The absolute magnitude in the $B$ band $M_B$ was calculated  using the  $g$ and $r$ magnitudes  and the Robert Lupton's  equations published on the SDSS web-page\footnote{\url{http://www.sdss3.org/dr8/algorithms/sdssUBVRITransform.php\#Lupton2005}}.  The foreground
Galactic extinction values from the NASA/IPAC Extragalactic Data
base (NED)\footnote{\url{http://ned.ipac.caltech.edu}} and K-corrections from \citet{Chilingarian2010} were taken into account.

The effective radius of the  galaxies $r_{eff}$   was obtained from the $r$-SDSS  estimations by  \citet{Reshetnikov2015} or from the Hyperleda database \citep{Leda}. The values  $r_{eff}$ in arcsecs are listed in Table~\ref{tab:data_info}.

\begin{figure*}
	\includegraphics[width=\linewidth]{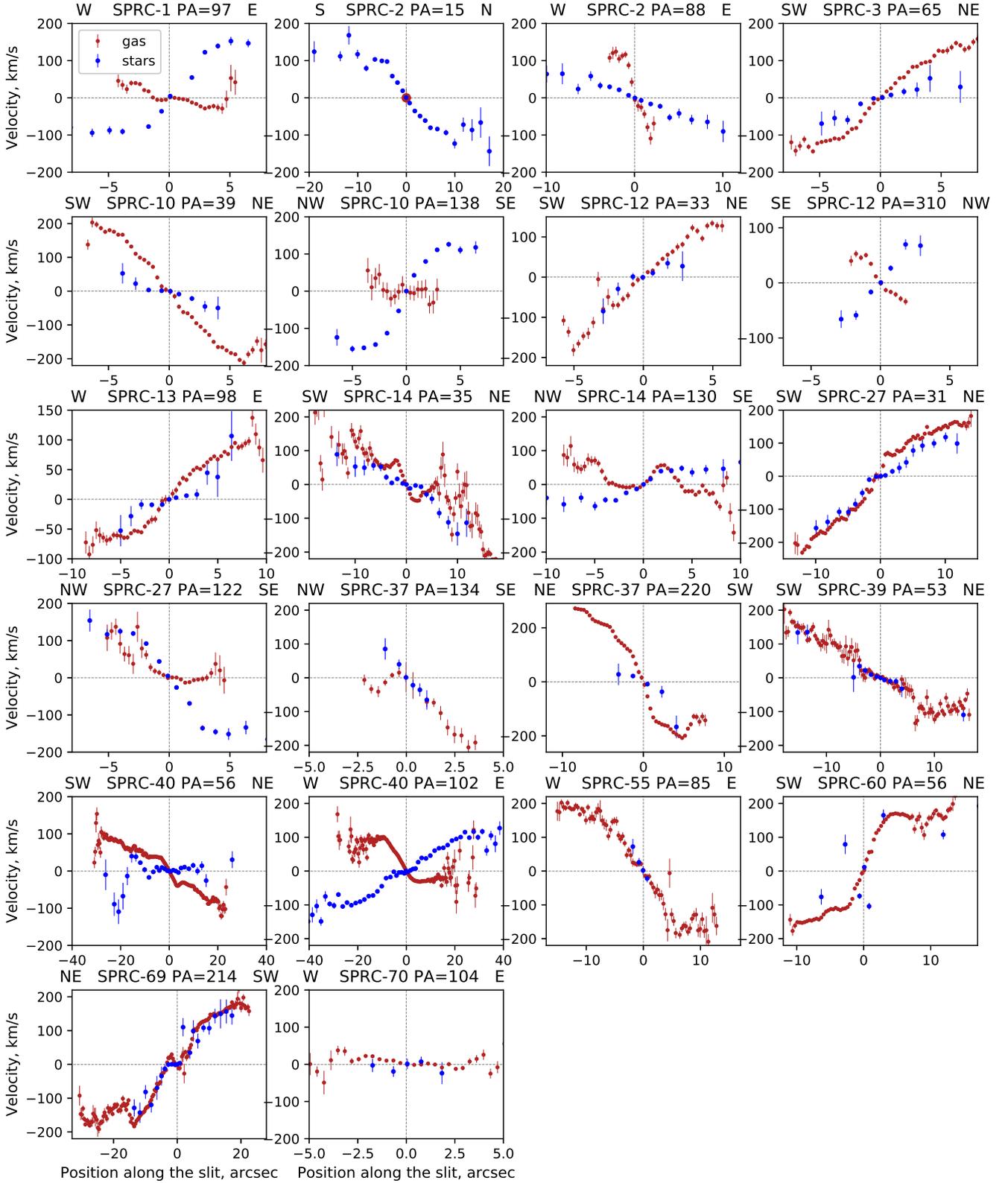}
	\caption{Line-of-sight velocities of  the stars (blue) and of the ionized gas in the \Ha+\NII\ emission lines (red) along the slits for each galaxy. The velocities are shown relative to the values corresponding to the galaxy centre.}
	\label{fig:vel}
\end{figure*}

\section{Gas and stellar velocities}
\label{sec:kin}

The detailed comparison between the rotation curve of the host galaxy and surrounding polar structure is a very important issue, because it allows one to reconstruct the 3D shape of the dark matter halo in PRGs. The examples of this reconstruction based on the 6-m telescope spectroscopic data in the case of SPRC-7 and SPRC-33 (NGC 4262) were already published by \citet{Khoperskov2014}. The stellar and ionized gas kinematics in  the observed  SPRC galaxies will be considered in detail in the forthcoming paper. In this Section, we only briefly describe the distributions of the line-of-sight velocities in the considered  sample. We focused our analysis on the kinematic  evidence of multi-spin components, because only 6 of 15 galaxies have been already kinematically confirmed as PRGs according to the previous studies (see below).

The line-of-sight velocities  of stars  were derived from fitting the observed spectra with a synthetic stellar population model as described in the previous Section. The ionized gas was detected in the residual spectra (after the \textsc{ULySS} model subtraction) in  all sample galaxies. It is distributed in polar rings mostly, while it is significantly fainter and less extended in the main discs of some galaxies. In the case of  SPRC-2 (PA$=15\degr$),  the  ionized gas emission was detected only in the galactic nuclear after integrating of the surrounding spectra.  The radial distributions of the line-of-sight velocities of the stellar and gaseous  components along each slit positions  are shown in Fig.~\ref{fig:vel}. 
The ionized gas velocities were derived using the single Gaussian fitting of the \Ha emission line, which was the brightest in the observed spectral region, together with the neighbouring \NII$\lambda6548,6583$ doublet. The line-of-sight velocity and Gaussian width were the shared parameters for the fitting of all these emission lines. We estimated the uncertainties of the measured velocities of the ionized gas by analysing a large number of synthetic spectra with a given signal-to-noise ratio, while the uncertainties of the stellar velocities were derived by \textsc{ULySS.}

In the case of a `classical' PRG having a gas-rich polar component around an early-type gas-free central galaxy, the  following kinematic  picture is expected:
\begin{enumerate}
    \item The same systemic velocity of the gaseous and stellar components.
    \item A large radial gradient of the gas velocities and zero gradient of the stellar line-of-sight velocities  along the major axis of the polar ring.   
    \item  A reverse situation, i.e., a large velocity gradient of stars and almost constant gas velocities along the host galaxy major axis.
\end{enumerate}

However, this  simple picture (constant velocities of stars  together with a strong rotation of gas and vice versa) is clearly presented in Fig.~\ref{fig:vel} only in SPRC-10 along $PA=138\degr$. If the polar ring is not strictly orthogonal to the central galaxy, the both components will demonstrate the velocity changing along the slit. However, their radial gradients will be significantly different: the larger value corresponds to the component rotating with line-of-nodes near the slit PA. The observed velocity distribution have become more complicated, when one detects the rotation of stars  in the host  galaxy and in the  polar ring simultaneously along  the same slit but on  different galactocentric distances. The SPRC-3 in Fig.~\ref{fig:vel} is a typical example. In the majority of the galaxies, both effects  described above are observed, i.e., the non-orthogonality of multi-spin components having  stars and gas. One of the most complex structures was observed in SPRC-14: a warped polar ring with a possible  collision between its gas clouds with the gas belonging to the host galaxy \citep[see also][]{sprc}.

In this paper, we analyse  the  long-slit spectra already used for the successful kinematic confirmation of large-scale polar rings in SPRC-10, SPRC-14, SPRC-39, and SPRC-60 \citep{sprc}. The polar ring in SPRC-69 was also confirmed in the same paper, but spectra in two cross-sections were obtained in the smaller  spectral range. In the present work, we obtained new spectroscopic data in the whole optical range. The brief description of the  kinematics of the rest sample galaxies is presented below.
\\
\textbf{SPRC-1: confirmed PRG}. The spectra taken along the host disc major axis demonstrate the flat rotation curve of the stellar component with a projected amplitude of $100-150\kms$, while the gas rotation is absent at the distances  $r<2''$,  where the slit crosses the brightest part of the edge-on polar ring (see \citealt{Reshetnikov2011} and the high-resolution image in the  ESO Press Release\footnote{\url{https://www.eso.org/public/images/eso9845c/}}). The  counter-rotation of the gas with an  amplitude of smaller than $40\kms$ along the line of sight might be related with a warp of the external part of a polar ring.  
\\
\textbf{SPRC-2: confirmed PRG}. The  extended emission of the ionizied gas is detected only along the polar ring which is non-orthogonal to the stellar disc. 
\\
\textbf{SPRC-3: confirmed PRG}. The rising rotation of the gas along the polar ring up to $\pm150\kms$ is observed. The  gradient of stellar velocities in the host disc is absent, while at $r>2''$ the stellar velocities  in the polar component  appear.
\\
\textbf{SPRC-12: confirmed PRG}. The central host disc is moderately inclined to the line of sight with the major axis  along $PA=310\degr$. The ionized gas in this  direction of the slit reveals the counter-rotation to the stars with a twice smaller amplitude. Together with a larger velocity gradient  along  $PA=33\degr$ (the photometrical major axis of the ring, see Fig.~\ref{fig:sdss}) the observed gas kimenatics corresponds to the inner warp of the polar ring. The rotation of the stellar population is also detected in the ring  along  $PA=33\degr$.
\\
\textbf{SPRC-13: confirmed PRG}. The velocity distribution is similar to that of SPRC-3: the strong ionized gas velocity gradient along the ring; the stellar velocities are absent near the centre (the host minor axis); the rise of the velocities  at $r>4''$ is related to  the stellar population in the ring. 
\\
\textbf{SPRC-27: confirmed PRG}. The kinematic behaviour of stars and gas similar to those of SPRC-13 is observed along the the polar ring  ($PA=31\degr$). It is interesting that along the host disc major axis ($PA=122\degr$) the gas belongs to the polar ring only in the SE side ($r>0$), while in the NW side of the disc, it possesses the ionized gas rotating in the same plane as stars. 
\\
\textbf{SPRC-37: confirmed PRG}. The kinematics along $PA=220\degr$ is typical of the polar ring major axis: a large ($\sim200\kms$) amplitude of gas rotation, the constant velocities  of the stellar component (the minor axis of the  host disc), the stellar velocities in the ring are also detected at $r\approx+5''$. The spectra along $PA=134\degr$  are too weak for a detailed analysis, however,the  rotation of the central disc is detected both in stars and gas.
\\
\textbf{SPRC-40 (NGC 5014): confirmed PRG}. A blue star-forming ring inclined at $\sim45\degr$ to the lenticular host  appears in the SDSS images. The HI velocity field demonstrates the ring rotation around the main galaxy and is connected with the extended gaseous tidal structure \citep{Noordermeer2005}. The spectra taken along the host galaxy major axis ($PA=102\degr$) reveal a significant asymmetry in the  gas rotation: the maximum velocities  in the Western side are about $100\kms$, while they do not exceed $40-50\kms$ in the Eastern side. The observed gas-stars counter-rotation  clearly indicates  that all the ionized gas is of external origin related to the inclined ring. The spectra taken at $PA=56\degr$ demonstrate a symmetric  gas rotation curve along the ring major axis. 
\\
\textbf{SPRC-55 and SPRC-70.} The spectra were taken only in a single slit position; both demonstrate a good agreement between the stellar and gas rotation, but the stellar kinematics was measured only near the nucleus. New spectroscopic data are necessary for  confirmation or rejection of polar rings in these candidates.

Therefore, together with the previous published results we have confirmed multi-spin structures in 13 of 15 candidates in the  sample.

\begin{figure*}
	\includegraphics[width=\linewidth]{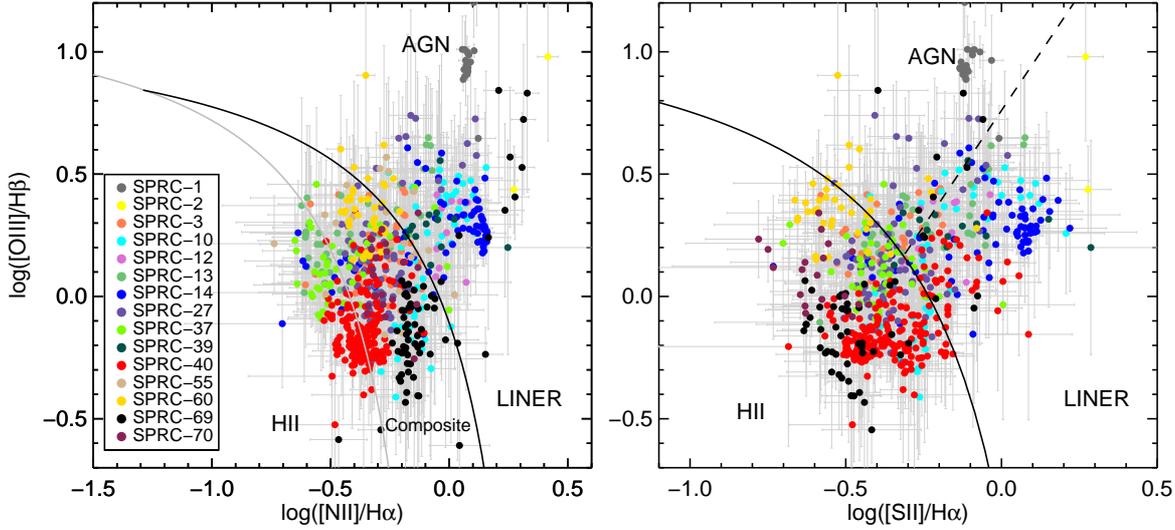}
	\caption{BPT-diagrams constructed for each spatial pixel in each  spectrum of the entire galaxy sample. Different symbol colours correspond to different galaxies. The black curve in both diagrams represents the ‘maximum
starburst line’ from \citet{Kewley2001}, and the grey curve from \citet{Kauffmann2003} in the left-hand diagram  separates the pure star-forming regions from those with a composite mechanism of the gas excitation. Dashed line from \citet{Kewley2006} in the right-hand panel separates the areas typically occupied by AGNs and LINERs (low-ionization narrow emission-line regions).}\label{fig:bpt_pix}
\end{figure*}

\begin{figure*}
	\includegraphics[width=\linewidth]{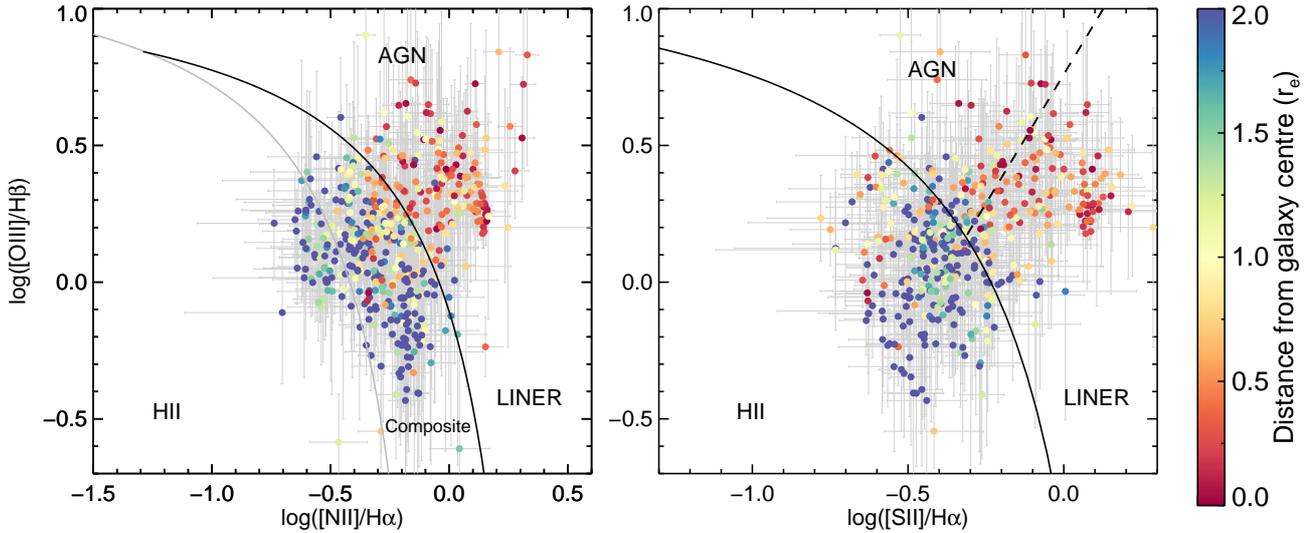}
	\caption{BPT-diagrams constructed for each spatial pixel in each obtained spectrum. Colours denote the distance from the galaxy centre normalized to its effective radius according to the scale presented. SPRC-1 and SPRC-40 were excluded from the analysis.}
	\label{fig:bpt_rad}
\end{figure*}

\begin{figure*}
	\includegraphics[width=\linewidth]{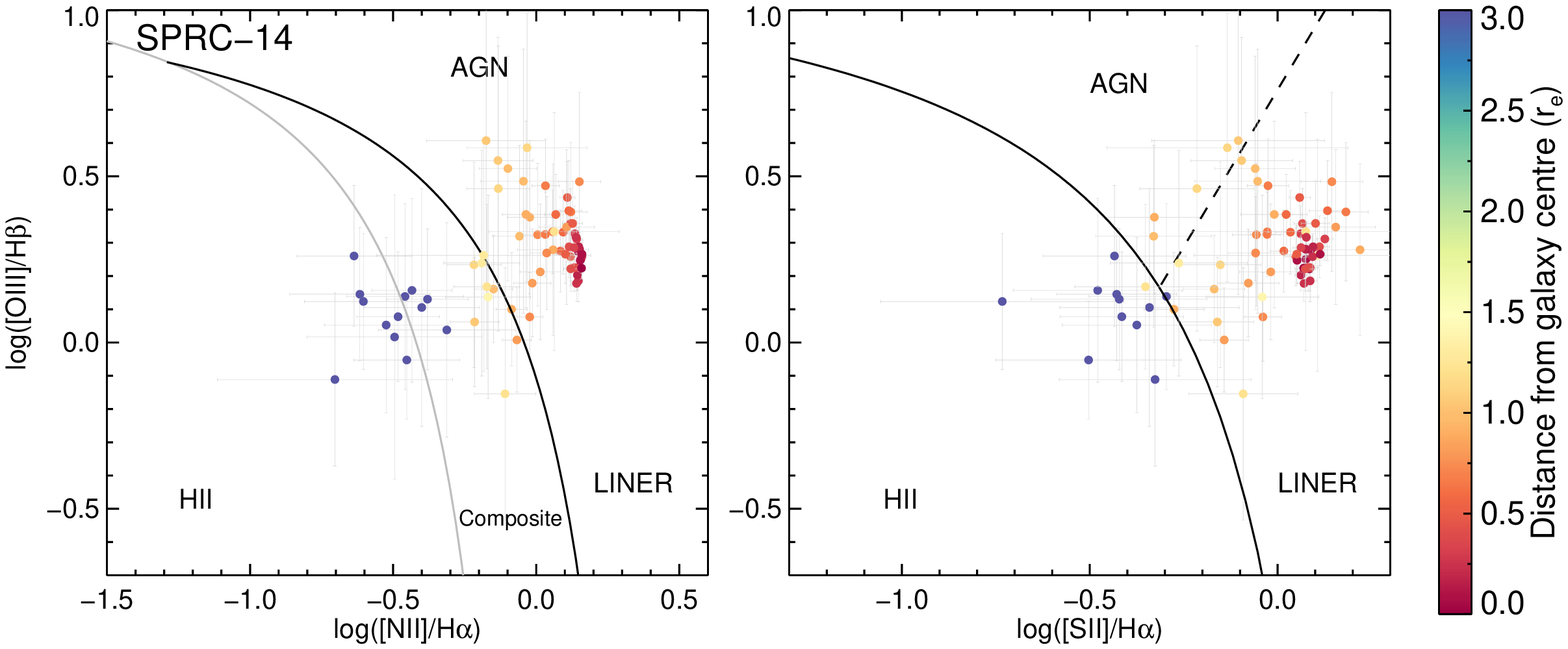}
    \includegraphics[width=\linewidth]{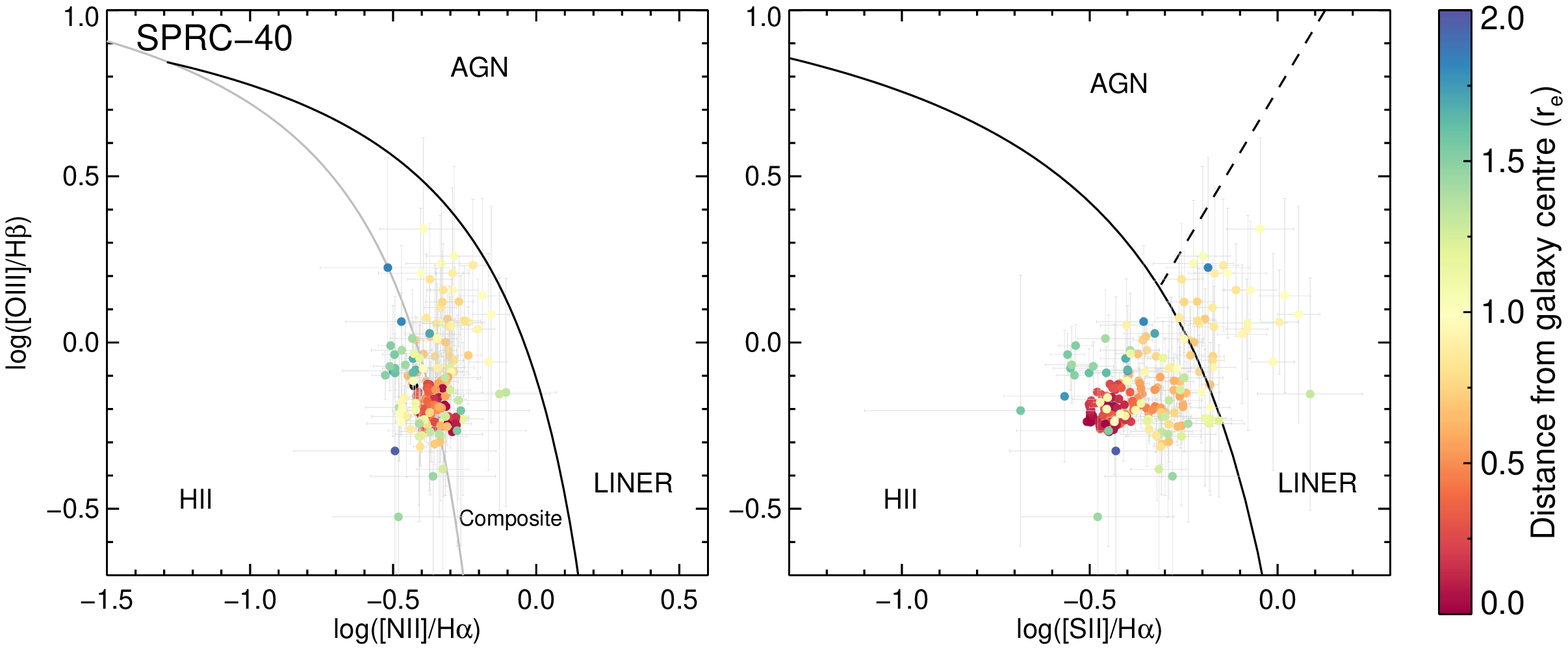}
	\caption{BPT-diagrams similar to those in Fig.~\ref{fig:bpt_rad} separately for SPRC-14 (top) and SPRC-40 (bottom).}
	\label{fig:bpt_14_40}
\end{figure*}

\section{Gas ionization state in polar rings}
\label{sec:ionization}

Ionization conditions of the gas were evaluated  via  the classical \OIIIHb\, versus \NIIHa\, and \SIIHa\, diagnostic diagrams \citep*[so-called BPT-diagrams, after][]{bpt}. Fig.~\ref{fig:bpt_pix} shows the BPT-diagrams for each spatial pixel in each spectrum for the observed galaxies. The black curve shows the \citet{Kewley2001} `maximum starburst' line  demarcating the OB-star  photoionization  from the area with domination of other mechanisms of the line excitations   (shocks, AGN, etc.). The grey line from \cite{Kauffmann2003} separates the areas of pure photoionization   from the regions with the composite mechanism of excitation -- both UV photons and shocks might be important there. As it follows from the diagrams, most of the observed galaxies demonstrate different mechanisms of ionization at different galactocentric distances, but the composite mechanism dominates in the \OIIIHb\, vs \NIIHa\, diagram, while 45 per cent of all points lies in the shock ionization domain in the \OIIIHb\, vs \SIIHa\, diagram.  The line ratio in the whole radial range  of SPRC-1 corresponds to the ionization by an Active Galactic Nucleus (AGN). 

In order to spatially separate the shock-dominated areas from the \HII regions, we coded each point on the BPT-diagrams in Fig.~\ref{fig:bpt_rad} according to the projected galactocentric distances along the slit normalized to the galaxy effective radius listed in Table~\ref{tab:data_info}. Two galaxies were excluded from this plot:  SPRC-1 (the only AGN galaxy) and  SPRC-40 (it is shown separately in Fig.~\ref{fig:bpt_14_40} and discussed below). 

One can observe that almost all points lying to the right of the `maximum starburst' line are located within one effective radius, while both photoionization and shocks are  observed at the outer parts of the polar rings. A noticeable gradient of the ionization state with the galactocentric radius  exists. 

Thus, we may conclude that shocks are observed close to the host galaxy, while its contribution is significantly lower in the outer regions of the polar disc. \cite{Freitas2012} obtained the same result for the AM 2020-504 galaxy. These findings are consistent with the \citet{Wakamatsu1993} hypothesis: shocks are generated, when gas on polar orbits crosses the potential well of a stellar disc. Indeed, the \citet{Wakamatsu1993} toy model says that stronger shocks might be generated in the interaction with denser disc regions, i.e., in the internal radii of a polar structure.  

A vivid example of the considered relation is the galaxy SPRC-14, its BPT diagrams are shown separately on the top panels in Fig.~\ref{fig:bpt_14_40}. The polar structure in this galaxy is well resolved; it has one of the largest angular sizes in the sample. We are able to study the gas excitation in the polar ring even along the slit with $PA=130\degr$ (the host galaxy major axis), because the ring is moderately inclined to the line-of-sight.  SPRC-14 contains a weak AGN  (of the LINER-type)  according to the BPT-diagrams. However, the shock ionization appears also at the intermediate  radial range (0.5--2.5$r_e$) between the nucleus and  outward regions in the star-forming polar ring. The observed gas kinematics provides evidence of the interaction between gaseous clouds on the polar orbits and in the host galaxy disc (see notes in Sec.~\ref{sec:kin}). It is possible that the ionization in the intermediate radii in SPRC-14 is caused not only by an intersection of polar gas clouds with a stellar disc as suggested by \citet{Wakamatsu1993} but also by a direct collision of clouds in both multi-spin components. The similar picture was observed, for instance, in Arp~212 galaxy \citep{Moiseev2008arp212}.

The bottom panels in Fig.~\ref{fig:bpt_14_40} show  an exception to the discussed relation:  the emission line ratios in the central region of SPRC-40 correspond to the \HII regions, whereas at  larger radii, the points lie in the `composite' or `shock' areas on the BPT diagrams. This radial changes of the ionization state is due to the specific morphology of SPRC-40 (NGC 5014) differing from other `classical' polar rings in our sample (Fig.~\ref{fig:sdss}): the low-contrast ring has a significantly smaller size than the stellar disc. Moreover, the ring is not orthogonal, but tilted on about $45\degr$ to the host galaxy. Along the disc major axis ($PA=102\degr$) in the central region we observed star formation in the accreted gas already settled to the main plain and counter-rotated to the stars (Sec.~\ref{sec:kin}). Whereas at distances of $8-12''$, the slit crosses the dust lane, that is obviously related to shocks generated by collision between the tilted ring and the disc. The angular scale in NGC 5014 (the closest galaxy in the sample) is 8--20 times better than that in other studied SPRC galaxies; it gives us a `zooming up view' on the process of the polar/inclined ring interaction with the disc in the circumnuclear region that is not resolved in other cases. 
Indeed, the SCORPIO-2 long-slit located along the  ring major axis ($PA=56\degr$) does not cross the disc-ring regions of intersection. The corresponded spectra demonstrate the \HII-like ionization up to  $r\approx1.5-2r_e$ (the blue and green points in Fig.~\ref{fig:bpt_14_40}, bottom).

\section{Gas metallicity and [O/H] gradient in polar rings}
\label{sec:metallicity}

Measurements of gas metallicities are complicated for PRGs because of the above-mentioned high contribution of shocks to the emission line excitation, while all the developed strong line emission methods for estimating oxygen abundance (as a gas metallicity indicator) were calibrated on the `normal' \HII regions. For further analysis, we excluded all points lying to the right of the `maximum starburst' line on the BPT-diagrams. After that, we were able to trace the metallicity gradients in 10 galaxies of our sample (in their polar rings) and to estimate the mean oxygen abundance derived from their stacked spectra collected  from  radial bins which   lie below the `maximum starburst' line on the BPT diagrams. The spectra of the selected regions were shifted to the central velocity of a galaxy before stacking.

\begin{figure}
	\includegraphics[width=\linewidth]{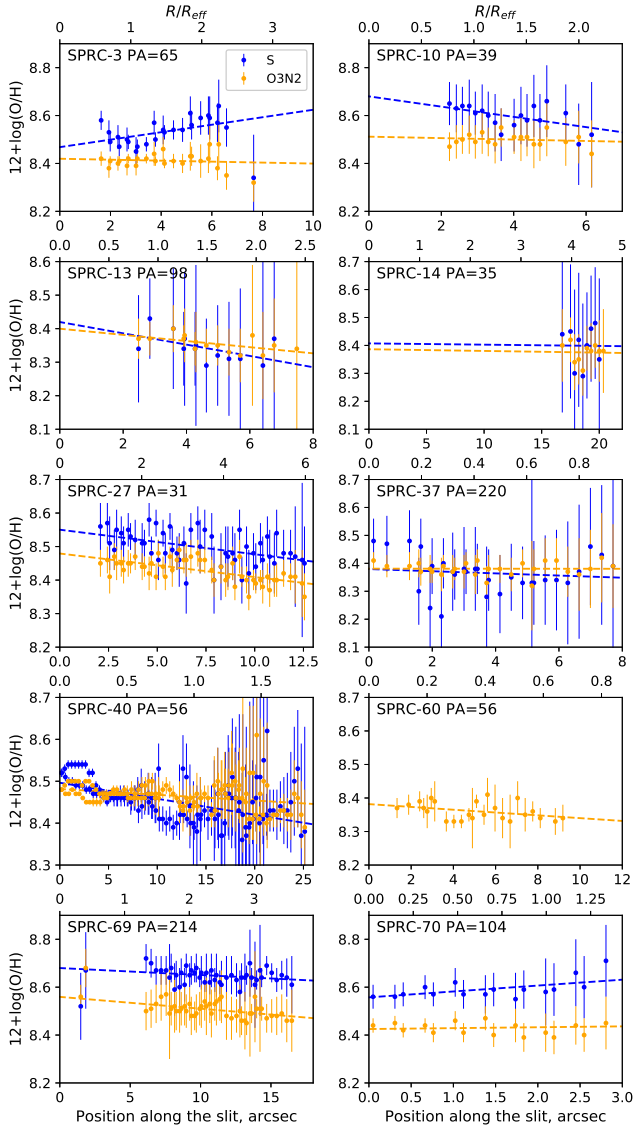}
	\caption{Radial metallicity distribution for several sample galaxies. The dashed lines mark the linear approximation of the [O/H] gradient for the estimations by S (blue) or O3N2 (yellow) methods. In SPRC-70, the plots correspond to the host galaxy major axes, in other galaxies -- to the major axis of the polar ring.}\label{fig:metgrad}
\end{figure}

	\begin{table*}
		\caption{Results of metallicity analysis}\label{tab:abund}
		\begin{small}
			\begin{tabular}{lccccc}
				\hline
				Galaxy  & M$_{B}$ (total) & M$_{B}$ (pol. disc) &  12+$\log$(O/H)$_\mathrm{O3N2}$ & 12+$\log$(O/H)$_\mathrm{S}$ & 12+$\log$(O/H)$_\mathrm{HCm}$\\
				\hline
				SPRC-1 & $-20.26 \pm 0.05$ & $-$ & $-$ & $-$ & $-$ \\
SPRC-2 & $-20.25 \pm 0.02$ & -18.37 & $-$ & $-$ & $-$ \\
SPRC-3 & $-18.57 \pm 0.06$ & -18.29 & $8.41\pm0.08$ & $8.50\pm0.05$ & $8.54\pm0.06$ \\
SPRC-10 & $-18.72 \pm 0.04$ & -15.78 & $8.51\pm0.08$ & $8.62\pm0.06$ & $8.70\pm0.11$ \\
SPRC-12 & $-18.70 \pm 0.06$ & -16.97 & $8.48\pm0.16$ & $8.63\pm0.24$ & $-$ \\
SPRC-13 & $-19.19 \pm 0.04$ & -17.70 & $8.36\pm0.08$ & $8.33\pm0.07$ & $8.54\pm0.15$ \\
SPRC-14 & $-19.74 \pm 0.02$ & -18.65 &$8.39\pm0.09$ & $8.42\pm0.09$ & $8.42\pm0.17$ \\
SPRC-27 & $-19.15 \pm 0.04$ & -18.08 & $8.43\pm0.08$ & $8.50\pm0.05$ & $8.62\pm0.08$ \\
SPRC-37 & $-19.52 \pm 0.04$ & -18.36 & $8.38\pm0.08$ & $8.36\pm0.06$ & $8.61\pm0.10$ \\
SPRC-39 & $-18.06 \pm 0.05$ & -16.71 & $-$ & $-$ & $-$ \\
SPRC-40 & $-17.30 \pm 0.02$ & $-$ & $8.47\pm0.08$ & $8.48\pm0.05$ & $8.62\pm0.05$ \\
SPRC-55 & $-19.67 \pm 0.09$ & -18.86 & $-$ & $-$ & $-$ \\
SPRC-60 & $-19.42 \pm 0.08$ & $-$ & $8.38\pm0.08$ & $-$ & $-$ \\
SPRC-69 & $-18.98 \pm 0.03$ & -17.28 & $8.51\pm0.08$ & $8.65\pm0.05$ & $8.70\pm0.07$ \\
SPRC-70 & $-18.83 \pm 0.12$ & -18.07 & $8.42\pm0.08$ & $8.54\pm0.05$ & $8.73\pm0.07$ \\
				\hline
			\end{tabular}
		\end{small}
	\end{table*}
	
We used three methods of oxygen abundance estimation: two `empirical' (calibration is based on the  metallicity of the \HII regions derived with the $T_e$-method): S \citep{Pilyugin2016} and O3N2 \citep{Marino2013} and one `theoretical' (calibration is  based on the photoionization models) HII-CHI-mistry \citep[HCm;][]{PerezMontero2014}. All the methods rely on the \OIIIHb\, and \NIIHa\, strong line ratios, while the S method also involves  [S~\textsc{ii}]/[N~\textsc{ii}], and the HCm uses \SIIHa\, and  I([O~\textsc{ii}])/I(H$\beta$)\, where available. 
The results are summarized in Table~\ref{tab:abund}, where the columns give: the galaxy name (according to the SPRC catalogue); its total absolute magnitude $M_B$ (measured in this work) and that of polar structures  only (calculated using the same equation as mentioned in Sec.~\ref{sec:obs} from  the corresponding $g$ and $r$ luminosities  obtained  by \citealt{Reshetnikov2015} from the 2D image  decomposition); the oxygen abundances $\mathrm{12+\log(O/H)}$  derived from the stacked spectra with three  methods mentioned above. 
Different methods give  different results for the same spectra that is a consequence of the well-known and still unsolved problem of discrepancy between different abundance calibrators \cite[see, e.g.,][]{Kewley2008}: the systematic offset between different methods might be up to 0.5 dex. Fortunately, in the case of the PRG sample, this difference is about 0.1-0.2 dex, i.e., the majority of the methods applied for the same galaxy agree within the uncertainties presented in Table~\ref{tab:abund}. Anyway, all the methods used reveal the sub-solar metallicity for the considered PRGs but $12+\log\mathrm{(O/H)}>8.3$ in each case.  
	
Fig.~\ref{fig:metgrad} shows the radial distribution of the oxygen abundance derived with the S and O3N2 methods  for ten galaxies from our sample. The observed metallicity gradients are almost negligible in most cases:  $\Delta(O/H) = -0.005 \div -0.07$ dex~$R_{eff}^{-1}$, that is much less than $ -0.1$ dex~$R_{eff}^{-1}$ typically observed in disc galaxies from the CALIFA \citep{Sanchez2014} or MaNGA \citep{Belfiore2017} surveys. Because of the shallow metallicity gradients, we may expect that the stacking procedure described before allows us to reliably estimate the mean oxygen abundance within the uncertainties  presented in Table~\ref{tab:abund}.

	\begin{figure*}
		\includegraphics[width=0.495\linewidth]{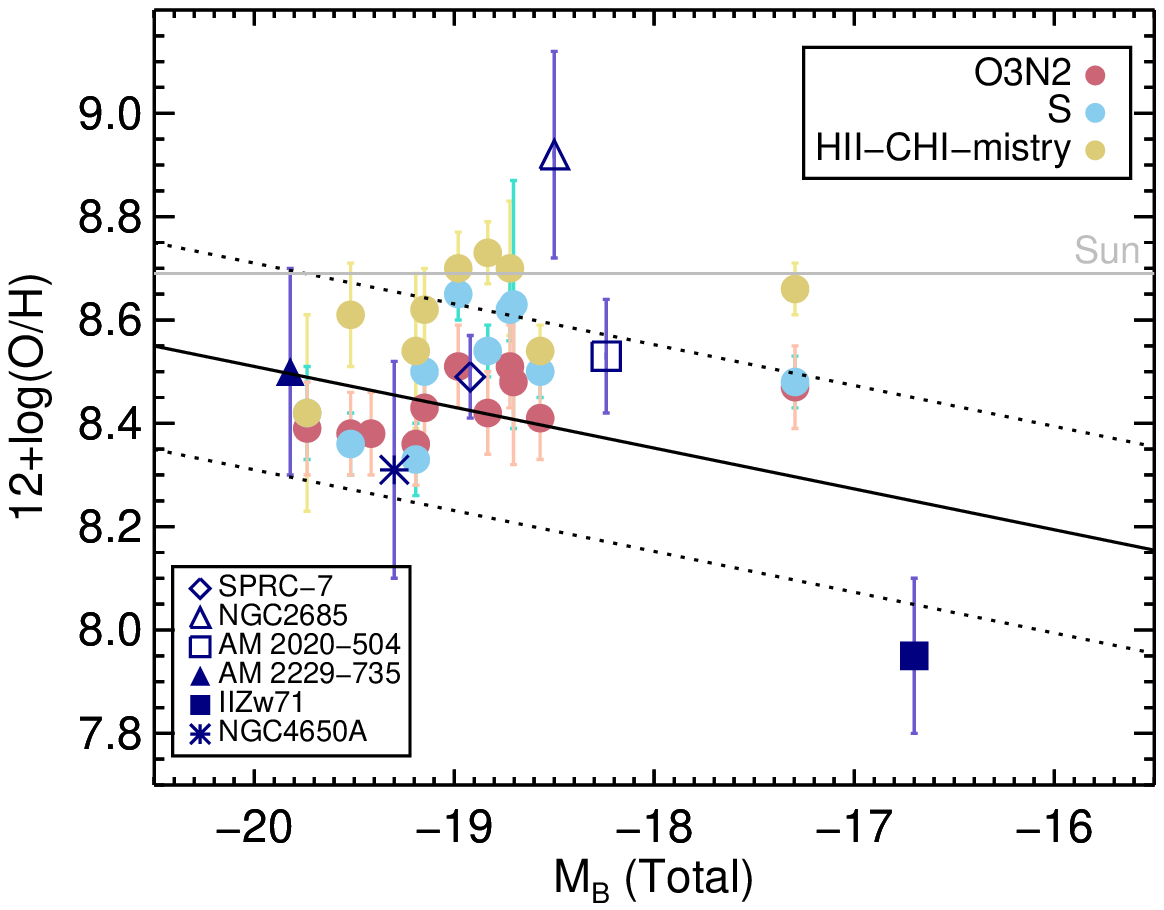}~\includegraphics[width=0.495\linewidth]{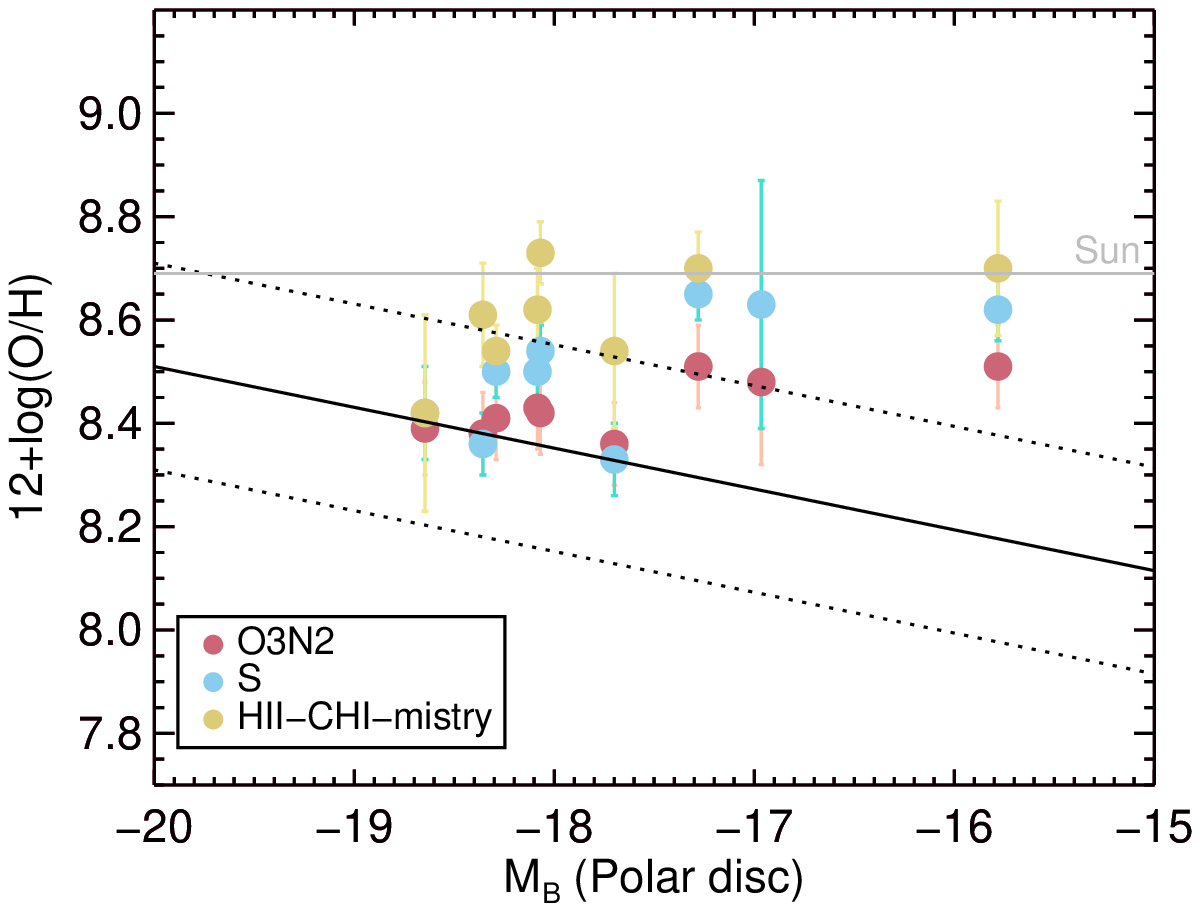}
		\caption{Luminosity--metallicity relation for our sample galaxies. All the points correspond to the total absolute magnitude of a galaxy in the left-hand panel and to the magnitude of the polar component only in the right-hand panel. The circles of different colours correspond to different calibrators used for metallicity estimation. The other symbols denote the values taken from the literature: SPRC-7 \citep{Brosch2010}, NGC2685 \citep{Eskridge1997}, AM 2020−504 \citep{Freitas2012}, AM 2229-735 \citep{Freitas2014}, IIZw71 \citep{PerezMontero2009}, and NGC4650A \citep{Spavone2010}. The black solid line shows the O/H -- $M_B$ relation for spiral galaxies \citep{Pilyugin2004}; the dashed lines correspond to its uncertainty ($\pm 0.2$~dex). }\label{fig:LZ}
	\end{figure*}

\section{Discussion}
\label{sec:discus}

It is  interesting to use the findings from the preceding Section (the sub-solar gas metallicity and the shallow [O/H] gradient) to test different evolutionary scenarios of the polar structure formation, especially the most intriguing one -- the cold gas accretion from cosmic filaments favoured, e.g., by \cite{Spavone2010}. 

In the case of a cold accretion scenario, the metallicity of polar rings is expected to be low -- $12+\mathrm{(O/H)} \sim 7.6-7.9$ according to the models by \citet{Snaith2012} for the prototype of the NGC4650A galaxy. These authors also predicted very shallow gas metallicity gradients ($-0.005 \div -0.01\ \mathrm{dex\ kpc^{-1}}$) in the polar components.  Despite this, we observe the oxygen abundance gradients to be very similar to their predicted values, we do not see any low-metallicity gas in our sample galaxies. The shallow metallicity gradients might point to the well mixing of gas during the evolution of the polar structure formed  via the `ordinary' tidal accretion  from a gas-rich companion. This is typical of interacting and merging galaxies \citep[see, e.g.,][]{Rupke2010a, Rupke2010b, Zasov2015}. Moreover, the relatively low metallicity of the accreted gas is expected, because the matter is  captured from the periphery of a donor  galaxy \citep{Bournaud2003}, where the gas usually has a  reduced metallicity in case of the negative radial [O/H] gradient. Therefore, the key question is: `how typical is the observed metallicity of a certain galaxy?'   

Fig.~\ref{fig:LZ}  shows the luminosity--metallicity relation for spiral galaxies taken from \citet{Pilyugin2004}. The points in the left-hand panel correspond to the oxygen abundance estimates (different colours mean different methods) and to the total absolute magnitude $M_B$ of the sample galaxies. In this figure, we also overlaid the measurements found in the literature (several unreliable points with reported errors of more than 0.5 dex have been excluded). As it follows from this figure, the oxygen abundance of PRGs is consistent with their total luminosity.
	 
If chemical abundance of the gas in polar rings is regulated mostly by evolution of the whole galaxy, then the total luminosity is responsible for  the physics in the luminosity--metallicity relation. On the other hand, if polar structures were formed by external processes like tidal accretion, then a large amount of metals in the gaseous disc were already presented in the donor galaxy. In this case, the metallicity of a polar ring should correlate with the luminosity of a donor galaxy (but not of a host galaxy), and  the luminosity of a polar ring will be a lower limit of the donor galaxy luminosity only.

The right-hand panel of Fig.~\ref{fig:LZ} shows the luminosity--metallicity relation for the observed galaxies, but using the absolute magnitude $M_B$ of polar rings only (results revealed from \citealt{Reshetnikov2015} and listed in Table~\ref{tab:abund}).
	
It can be clearly seen from Fig.~\ref{fig:LZ} (especially from the right-hand panel) that the metallicity of polar rings in the galaxies from our sample almost do not depend on the galaxy (or polar ring) luminosity. This fact might be easily explained in the frameworks of the tidal accretion scenario for the polar component formation. The chemical abundance of a polar ring in that case correspond to the values typical of donor galaxies but only a fraction of their gas was accreted onto the host galaxy of the PRG system and built the polar ring. Hence, the mass of gas and the luminosity of the polar ring could be significantly lower for their metallicities.
	
Summing up, we may rule out the cold accretion scenario  at least for the present sample of PRGs. The tidal accretion from a gas-rich donor or  major merger events are more favourable mechanisms for formation of the studied multi-spin structures. Therefore, the answer to the question raised in Sec.~\ref{sec:intro} is the following:  galaxies like NGC~4650A (considered as a `cold accretion prototype') are relatively rare  in the  present-day Universe. Extending of the sample of PRGs with spatially-resolved spectroscopic data, including 3D-spectral  surveys,  should check and quantify this conclusion.

\section{Summary}
	\label{sec:sum}

We carried out optical spectroscopic observations of 15 PRGs from the SPRC catalogue using the SCORPIO-2  long-slit spectrograph at the SAO RAS 6-m telescope. The ionized gas and stellar kinematics, the ionization conditions in polar rings, and the gas metallicity distribution were analysed. The obtained  results twice increase the number of kinematically-confirmed polar structures among SPRC galaxies and three times -- the number of PRGs with available measurements of chemical abundance. The most interesting findings are the following:

\begin{itemize}
		\item Shock waves are important in gas excitation in the inner regions of polar rings, while the relative contribution of photoionization by young stars grows at the outskirts of polar structures. This radial change of ionization sources is in a good agreement with the \citet{Wakamatsu1993} model, implying the collision between the polar gas and host stellar disc. Moreover, the direct collision between gaseous clouds in both multi-spin components is also possible.
		\item All 11 galaxies from our sample with the measured oxygen abundances have the sub-solar gas metallicity $Z=0.5-0.8 Z\odot$ that almost does not depend on the  luminosity of the galaxy or its polar structure. This result is independent of the accepted method  of oxygen abundance estimation (O3N2, S, or HCm).
		\item The radial distribution of metallicity is flat; a shallow or no metallicity gradient is observed. 
	\end{itemize}
	
Based on the above results, we ruled out a scenario of polar ring formation due to the cold accretion from cosmic filaments for the considered sample of polar-ring galaxies.

\section*{Acknowledgements}
This study was supported by the Russian Science Foundation, project no. 17-12-01335 `Ionized gas in galaxy discs and beyond the optical radius' and based on observations conducted with the 6-m telescope of the Special Astrophysical Observatory of the Russian Academy of Sciences carried out with the financial support of the Ministry of Science and Higher Education of the Russian Federation. 
We thank Vladimir Reshetnikov for kindly providing the data on photometry of SPRC galaxies and his comments about the metallicity gradient origin,  Dmitri Oparin and Roman Uklein for their assistance in the SCORPIO-2 observations, and especially Victor Afanasiev for his great contribution to spectroscopy at the 6-m telescope.
We thank the anonymous reviewer for the useful comments.
This research has made use of the NASA/IPAC Extragalactic Database (NED) which is operated by the Jet Propulsion Laboratory, California Institute of Technology under contract with the National Aeronautics and Space Administration. We acknowledge the usage of the HyperLeda database (http://leda.univ-lyon1.fr)	

Funding for the Sloan Digital Sky Survey IV has been provided by the Alfred P. Sloan Foundation, the U.S. Department of Energy Office of Science, and the Participating Institutions. SDSS-IV acknowledges
support and resources from the Center for High-Performance Computing at
the University of Utah. The SDSS web site is www.sdss.org.


\label{lastpage}

\end{document}